\documentstyle[11pt,psfig]{article}
\newcommand{\RR}{\rangle \rangle}
\newcommand{\LL}{\langle \langle}
\begin{document}
\renewcommand{\thepage}{ }
\begin{titlepage}
\title{
\hfill {\normalsize LPENSL-Th-04/99}
\vspace{1.5cm}
{\center \bf Spin glass behavior upon diluting frustrated
magnets\\and spin liquids: a Bethe-Peierls
treatment\thanks{This work is partly supported by TMR
grant FMRX-CT96-0012 (EC), CNRS (France) and MNERT grant AC-98-20511 (France)}
}
}
\author{
R. M\'elin$^{(1)}$ and S. Peysson$^{(2)}$\\
{}\\
{$^{(1)}$ Centre de Recherches sur les Tr\`es basses
temp\'eratures (CRTBT)\thanks{U.P.R. 5001 du CNRS,
Laboratoire conventionn\'e avec l'Universit\'e Joseph Fourier
}}\\
{CNRS, BP 166X, 38042 Grenoble Cedex, France}\\
{}\\
{$^{(2)}$ Laboratoire de Physique\thanks{U.M.R. CNRS 5672},
Ecole Normale Sup\'erieure de Lyon}\\
{46, All\'ee d'Italie, 69364 Lyon Cedex 07, France}\\
{}\\
}
\date{}
\maketitle
\begin{abstract}
\normalsize
A Bethe-Peierls treatment to dilution in frustrated
magnets and spin liquids is given.
A spin glass phase is present at low
temperatures and close to the percolation point
as soon as frustration
takes a finite value in the dilute magnet
model;
the spin glass phase is reentrant inside the
ferromagnetic phase.
An extension of the model is given,
in which the spin glass~/~ferromagnet
phase boundary is shown not to reenter inside the ferromagnetic phase
asymptotically close to the tricritical point whereas it
has a turning point at lower temperatures.
We conjecture similar phase diagrams to
exist in finite dimensional models not
constraint by a Nishimori's line.
We increase frustration to study the effect of dilution in
a spin liquid state.
This provides a ``minimal'' ordering by disorder
from an Ising paramagnet to an Ising spin glass.
\end{abstract}
\end{titlepage}
\newpage
\renewcommand{\thepage}{\arabic{page}}
\setcounter{page}{1}
\baselineskip=17pt plus 0.2pt minus 0.1pt

\newpage



\section{Introduction}
When non magnetic sites are diluted in an unfrustrated
ferromagnet
with a probability $\mu$, the transition temperature
is reduced and vanishes at the percolation threshold
$1-\mu_{P,0}$.
In such non
frustrated systems, two phases only exist: a low
temperature ferromagnetic phase above the percolation
threshold, and
a paramagnetic phase. 
If the temperature is decreased at
the percolation threshold,
the dynamics becomes slower
because large-scale droplet-like objects of size
$\xi_T$  form, with $\ln{\xi_T} \sim J/T$~\cite{Coniglio}.
These objects have energy barriers scaling like the
logarithm of their volume~\cite{Rammal,Henley}.
This results
in a slow dynamics and interrupted aging~\cite{RM-tree}
({\sl i.e.} with a finite relaxation
time~\cite{Bouchaud}).
This shows that despite the absence of frustration, the
simplest models of dilute magnets already have a phenomenology
close to the one of spin glasses, even though freezing in
these systems is a cross-over due to an increasing correlation
length becoming of order of the system size.
This indicates that some perturbations
of these unfrustrated systems may drive them to a true spin glass
phase, which we show in the present article by
studying the thermodynamics of a particular model.

In dilute magnet
compounds, such as Eu$_x$Sr$_{1-x}$S~\cite{EuSrS,note,exp},
a low temperature
spin glass phase appears close to the percolation
threshold. The main features of the phase diagram are:
\begin{itemize}
\item[(i)] As the dilution $\mu$ is increased from the pure system
with $\mu=0$, the ordering temperature decreases.

\item[(ii)] A tricritical point exists at a dilution $\mu_t$
and temperature $T_t$, with $1-\mu_t$ of order
of the percolation threshold $1-\mu_{P,0}$ in the absence
of frustration. At this tricritical point,
the ferromagnetic, paramagnetic and spin glass phases meet.

\item[(iii)] As dilution is increased from $\mu_t$,
the spin glass transition temperature decreases from
$T_t$ at the tricritical point to zero at the percolation
threshold $1-\mu_P$, with $1-\mu_P$ the percolation
threshold of the system with frustration, smaller than
the percolation threshold $1-\mu_{P,0}$ in the absence
of frustration.

\item[(iv)] The spin glass phase is reentrant inside the ferromagnetic
phase.
\end{itemize}
One purpose of the present article is to show that these qualitative
features of the phase diagram can be reproduced in a model
that combines dilution and short range
frustration. 
This model does not consist in a detailed microscopic
modeling of Eu$_x$Sr$_{1-x}$S, but rather contains
the generic ingredients entering the physics of these systems
(dilution and short range frustration)~\cite{Rammal-Souletie}.
Even though this treatment relies on
a specific lattice topology (a tree structure),
Bethe-Peierls phase diagrams are equivalent
to mean field phase diagrams while the
Bethe-Peierls method is powerful enough
to give an exact answer to the issue of reentrance.
The resulting phase diagrams are therefore not
expected to be specific to our treatment
but are generic features of the coupling
Hamiltonian.

The article is organized as follows.
The model is
given in section~\ref{sec:model}.
The paramagnetic phase boundary is solved in
section~\ref{sec:recur-rel}.
We study in section~\ref{sec:reentrance} the
spin glass~/~ferromagnet phase boundary and
show that it is reentrant. The issue of reentrance
is non trivial because Nishimori's
argument~\cite{Nishimori,Kitatani,Georges-LeDoussal,LeDoussal-Harris,Horiguchi}
does not
hold in our model, part of the exchanges
being frozen.
An extension
of the model is given in which the spin glass~/~ferromagnet
phase boundary is shown
not to reenter in the ferromagnetic phase
asymptotically close to the tricritical point,
whereas it has a turning point at a lower temperature.
We expect this unusual type of phase diagram
to be a generic feature of
models combining disordered and frozen
exchanges, and may be obtained in finite
dimensional models also.
Finally, we study in section \ref{sec:dil-spinliquid}
the effect of diluting a spin liquid state,
in which case ordering by disorder generates
a transition from an Ising paramagnet to
an Ising spin glass. This ordering by disorder
mechanism is ``minimal'' because spins
have an Ising symmetry only.

\section{The model}
\label{sec:model}

\subsection{The dilute magnet model and its generalization}
\begin{figure}
\centerline{\psfig{file=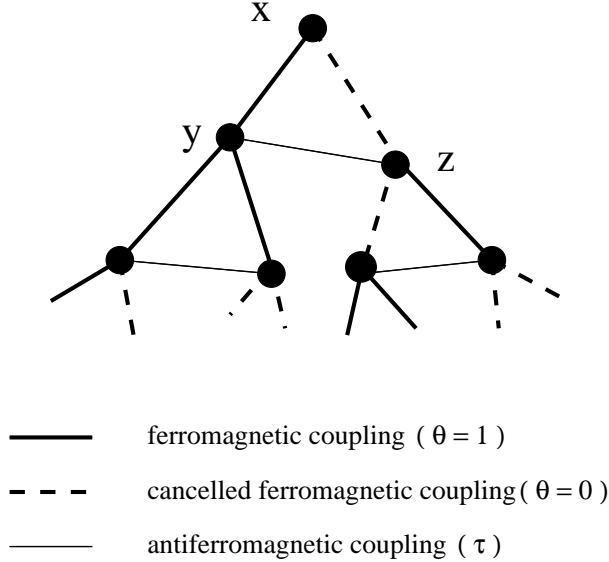,width=8cm}}
\caption{The Husimi cactus-like structure model
of dilute magnet with frustration. The
structure with a top spin $x$ is
obtained by gluing the two structures
with top spins $y$ and $z$. The top spin is
the highest one in the hierarchy (at site $x$
on this figure).
The ferromagnetic bonds of the tree are canceled
with a probability $\mu$ and a fixed antiferromagnetic
coupling $\tau$ is added.
}
\label{fig:husimi}
\end{figure}

In a
Bethe-Peierls calculation, only the properties of the ``top''
spin (the highest one in the hierarchy; see
Fig.~\ref{fig:husimi})
are considered,
and the thermodynamic limit is obtained by growing
the number of generations to
infinity~\cite{Bethe,Peierls,Baxter}. The top
spin fixed point magnetization distributions
$P^*(m)$ are of three types:
(i) {\it paramagnetic phase}: $P^*(m) = \delta(m)$;
(ii) {\it spin glass phase}:
$P^*(m)$ is even;
(iii) {\it ferromagnetic phase}: $P^*(m)$ has a finite first moment.
The transitions between the
phases are of a mean field
type~\cite{Bethe,Peierls,Baxter}. The phase diagram
of the $\pm J$ model on the Bethe
lattice~\cite{spinglass1,spinglass2,spinglass3,spinglass4,
spinglass5,Carlson,spinglass6}
is very similar to the one of the Sherrington-Kirkpatrick
model~\cite{SK},
and the Bethe-Peierls treatment allows a correct
description~\cite{spinglass6}
of the Almeida-Thouless line~\cite{Almeida-Thouless}.
This shows that a Bethe-Peierls
treatment succeeds in
reproducing  the mean field
phase diagram of spin glass models. 

We consider a model in which the ferromagnetic bonds
$J$ of the Cayley tree are canceled with
a probability $\mu$. Frustration is added by
completing the triangles, with frozen antiferromagnetic
bonds $\tau$ forming a Husimi cactus-like
structure~\cite{Thorpe} (see Fig.~\ref{fig:husimi}).
We use the binary variables $\theta_i=0,1$, with
$J_i = \theta_i J$. 
The temperature is expressed in units of $J$.
The Hamiltonian is
\begin{equation}
H = - \sum_{\langle i,j \rangle}
\theta_{i,j} \sigma_i \sigma_j
+ \tau \sum_{\langle i,j \rangle'}
\sigma_i \sigma_j
,
\end{equation}
where $\langle i,j \rangle$ denotes the bonds of the
tree structure and $\langle i,j \rangle'$ the
next nearest neighbor pairs of sites in the
same generation.
The distribution of the $\theta$-variable is 
\begin{equation}
\label{eq:bondinitial}
p(\theta) = (1-\mu) \delta(\theta-1) + \mu \delta(\theta)
.
\end{equation}
The specificity that some bonds
are frozen in this model has drastic
consequences on the shape of the
phase diagram, as we will show.

The Husimi cactus
structure allows the introduction
of a {\sl local} frustration resulting from
next-nearest-neighbor interactions,
and can be used to mimic
the effects of local antiferromagnetic interactions
in dilute compounds.
Chandra and Dou\c{c}ot~\cite{Doucot} considered a
frustrated spin model on a regular Husimi cactus structure, and
studied ordering by disorder
in the spin liquid state in the Bethe-Peierls limit
(see also~\cite{Quantum} for a study of
the effect of quantum fluctuations).
These authors considered a non disordered
model
in which a spin glass phase cannot
exist~\cite{Doucot}.
In our model with randomness, we show
the stability of a spin glass solution
in some regions of the phase diagram.

We consider bond instead of site percolation
because the site percolation threshold of the Husimi
cactus structure would be
equal to the one of the tree structure, independent
of the additional bonds $\tau$. The bond percolation model
is therefore better suited for modeling dilute compounds~\cite{EuSrS},
since the bond percolation threshold of the
structure without frustration (a tree structure) is $1-\mu_{P,0}=1/2$,
larger than the bond percolation threshold
$1-\mu_P =1-1/\sqrt{2}$
of the structure with frustration (the Husimi cactus structure 
shown on Fig.~\ref{fig:husimi}).

For the sake of generality, we not only consider
a dilute magnet model with $\theta=0,1$,
but extend the bond distribution~(\ref{eq:bondinitial}) to
incorporate possible antiferromagnetic bonds
on the tree structure. 
The distribution of the bond variables $\theta$ is
\begin{equation}
\label{eq:bond}
p(\theta) = (1-\lambda)(1-\mu) \delta( \theta - 1)
+ \mu \delta(\theta)
+ \lambda (1-\mu) \delta(\theta + 1)
,
\end{equation}
while the additional
antiferromagnetic bonds $\tau$ are frozen.
This model interpolates between the dilute model
with a short-range frustration $\tau$ ($\lambda=0$),
and the $\pm J$ model ($\mu=0$ and $\tau=0$).

\subsection{Absence of a Nishimori line argument}

Our Hamiltonian is formally invariant
under local gauge transformations
$\sigma_i \rightarrow \epsilon_i \sigma_i$,
$J_{i,j} \rightarrow \epsilon_i \epsilon_j \sigma_i
\sigma_j$, with $\epsilon_i=\pm 1$~\cite{Toulouse}.
In some spin glass models (such as the $\pm J$ model),
gauge invariance 
provides strong constraints on the phase
diagram. The internal energy can be calculated
exactly on Nishimori's line by expanding the
average energy over the gauge group~\cite{Nishimori}.
This line crosses the phase boundary at the tricritical
point~\cite{Nishimori,Kitatani,Georges-LeDoussal,LeDoussal-Harris}.
Moreover, spin correlations can be related to gauge
variable correlations, with the consequence that
the frontier between the ferromagnetic and
spin glass phases is either vertical or reentrant~\cite{Nishimori}.
In our model, gauge
invariance
is useless for the following reason. One can define
a local distribution of bond variables
$P_{i,j}(J)$, being (\ref{eq:bond}) on the tree
bonds, and $\delta(J+\tau)$ on the antiferromagnetic
bonds $\tau$. Nishimori's line is defined by
$
\beta_N = \frac{1}{2}
\ln{\left(\frac{1-\lambda}{\lambda} \right)}
$~\cite{Horiguchi}, and
$\beta_N = + \infty$.
The first equality originates from the
tree bond variables and the second 
from the frozen antiferromagnetic bonds $\tau$.
The two equalities can be formally met
if $\lambda=0$, in which case Nishimori line
is $\beta_N = + \infty$. However, this does
not make predictions
on the phase diagram possible
even in the case $\lambda=0$~\cite{Note}
because
one does not expect to be able
to describe finite temperature spin glass properties
in terms of the ground
state only (the only state
selected if $\beta_N=+\infty$).
Nishimori's
argument can therefore not be made in this model, and
the question of reentrance (item (iv)
in the introductory section) cannot be answered
on the basis of Nishimori's line while
Bethe-Peierls calculations are powerful
enough to allow the derivation of
exact results.
We show that the spin glass~/~paramagnet boundary is
reentrant in the dilution model
($\lambda=0$) in the $(\mu,T)$ plane.
In the $\pm J$ model with the additional
coupling $\tau$ ($\mu=0$),
and in the $(\lambda,T)$ plane,
we show that the spin glass~/~ferromagnet phase
boundary is not reentrant asymptotically close to the tricritical
point, whereas it has a turning point at lower
temperatures. This behavior, richer than
in usual spin glass models, is to our opinion a
generic feature of Hamiltonians combining
disorder and frozen bonds. We conjecture
the existence of finite dimensional models
with a similar phase diagram.

\section{Recursion relations}
\label{sec:recur-rel}
We now derive the recursion of the
top-site magnetization when cacti are glued 
as shown on Fig.~\ref{fig:husimi}. We denote
by $m_x$ the magnetization at site $x$, and 
$m_y$ and $m_z$ the magnetizations at the
descendant sites $y$ and $z$.
The derivation of the recursion relations in our dilute
magnet model is similar to the case of the $\pm J$ model
(see Ref.~\cite{Carlson}). 

\subsection{Recursions and the paramagnet phase boundary}
\label{sec:recursion}
Following
Ref.~\cite{Carlson}, we denote by ${\cal Z}^{(\pm)}_x$ the
conditional partition function with the spin
at site $x$ frozen in the direction $\pm$. The
magnetization at site $x$ is
$
m_x = ({\cal Z}^{(+)}_x - {\cal Z}^{(-)}_x)
/({\cal Z}^{(+)}_x + {\cal Z}^{(-)}_x)
.
$
The partition functions ${\cal Z}^{(\pm)}_x$ are related
to the partition functions ${\cal Z}^{(\pm)}_{y,z}$
of the descendant sites according to
\begin{equation}
\label{eq:Z}
{\cal Z}^{(\sigma_x)}_x = 
\sum_{\sigma_y,\sigma_z}
W^B_{\theta_y,\theta_z}(\sigma_x|\sigma_y,\sigma_z)
{\cal Z}_y^{(\sigma_y)}
{\cal Z}_z^{(\sigma_z)}
,
\end{equation}
with the Boltzmann weight factor
\begin{equation}
\label{eq:WB}
W^B_{\theta_y,\theta_z}(\sigma_x|\sigma_y,\sigma_z)
= \exp{\left( \beta (  \theta_y \sigma_x \sigma_y
+  \theta_z \sigma_x \sigma_z ) \right)}
\exp{\left( -\beta \tau \sigma_y \sigma_z \right)}
,
\end{equation}
and $\theta=0, \pm 1$.
We next trace over the spins at sites $y$
and $z$ in Eq.~\ref{eq:Z} to obtain
\begin{equation}
\label{eq:recursion}
m_x = f(m_y,m_z|\theta_y,\theta_z) =
p \frac{ m_y (\theta_y - u \theta_z)
+ m_z (\theta_z - u \theta_y)}
{1 - u p^2 \theta_y \theta_z
+ m_y m_z (p^2 \theta_y \theta_z -u)}
,
\end{equation}
with $p= \tanh{(\beta J)}$ and $u = \tanh{(\beta \tau)}$.
The recursion of the magnetization distribution is
\begin{equation}
\label{eq:PM}
P_{n+1}(m_x) = \int d m_x d m_y \sum_{\theta_y,\theta_z}
p(\theta_y) p(\theta_z) P_n(m_y) P_n(m_z)
\delta\left(m_x - f(m_y,m_z|\theta_y,\theta_z) \right)
,
\end{equation}
with $p(\theta)$ the distribution of bond variables
(\ref{eq:bond}), and $P_n$ the magnetization distribution
of the top spin with $n$ levels of hierarchy. We denote by
$\LL m^k \RR_n$ the moment of order $k$ of $P_n(m)$.

We now parametrize the tricritical line, where the three
phases (paramagnetic, ferromagnetic and spin glass) meet.
The meeting point of these phases is
a {\sl line} in the parameter space $(\lambda, \mu ,T)$.
If $\lambda$ [$\mu$] is fixed and the phase diagram is
considered in the $(\mu,T)$ [$(\lambda,T)$] plane,
the three phases meet in a tricritical {\sl point}.
Let us first consider the stability of the paramagnetic
solution with respect to perturbations in the first
moment. To lowest order the recursion of the first moment
is
$
\label{eq:recur-m}
\LL m \RR_{n+1} = 2 p \LL G_{y,z} \RR \LL 
m \RR_n
$, with 
$
G_{y,z} = (\theta_y - u \theta_z)/(1 - u p^2
\theta_y \theta_z)
$.
The disorder average of $G$ is understood as
$
\LL G_{y,z} \RR = \sum_{\theta_y,\theta_z}
p(\theta_y) p(\theta_z) G_{y,z}
$.
The paramagnetic solution is stable with respect to
perturbations in the first moment if
$2 p \LL G_{y,z} \RR < 1$. A similar reasoning
shows that the paramagnetic solution is stable with
respect to perturbations in the second moment
if $2 p^2 \LL G_{y,z}^2 \RR <1$.
To summarize, the tricritical line is defined by
\begin{equation}
\label{eq:tricritical}
2 p \LL G_{y,z} \RR = 1 \mbox{ , and }
2 p^2 \LL G_{y,z}^2 \RR = 1
.
\end{equation}

\subsection{Limiting cases}
\label{sec:limiting}
\begin{figure}
\centerline{\psfig{file=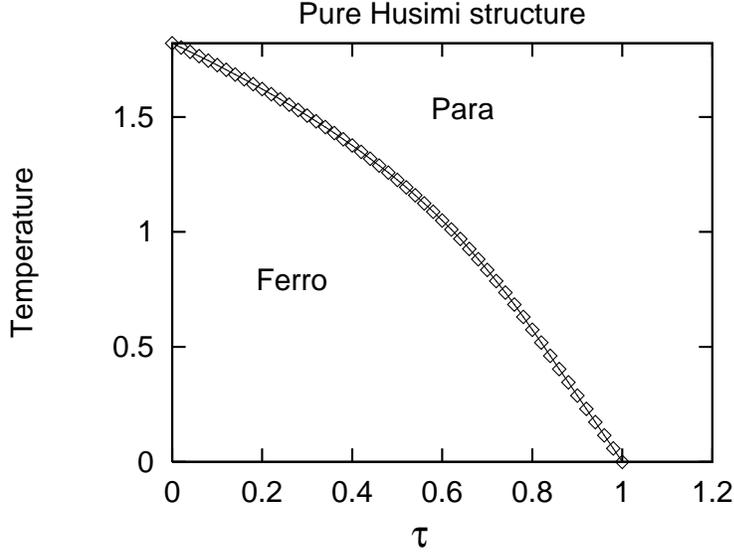,height=8cm}}
\caption{Phase diagram of the pure Husimi cactus system
(all the coupling $\theta$ being unity) as a function
of the frustration $\tau$. A low-temperature
ferromagnetic phase is present if $\tau<1$. If $\tau>1$, the system is
a spin liquid (it does not order even at zero temperature).}
\label{fig:tau}
\end{figure}
\noindent
{\sl The pure system:}
Let us consider the recursion (\ref{eq:recursion})
in the pure system limit in which the variables $\theta$ are
all equal to unity. 
This amounts to specializing the distribution (\ref{eq:bond})
to the case $\lambda=\mu=1$
while keeping finite the local frustration $\tau$.
The only possible phases are ferromagnetic and
paramagnetic.
The recursion of the magnetization is
$
m_{n+1} = 2 p (1-u) m_n/
(1 - p u^2 + (p^2 - u)m_n^2)
.
$
The paramagnetic phase is stable against ferromagnetic
fluctuations if
$
2 p (1-u)/(1 - p u^2) < 1
$.
The phase diagram is shown on Fig.~\ref{fig:tau} as 
a function of the frustration $\tau$ with a spin
liquid phase if $\tau > 1$.
When considering in
the following the frustrated magnet model, we assume
$\tau < 1$, in which case the pure system has
an ordered phase at low temperature. Diluting
the spin liquid state with $\tau>1$
is examined in section~\ref{sec:dil-spinliquid}.

\begin{figure}
\centerline{\psfig{file=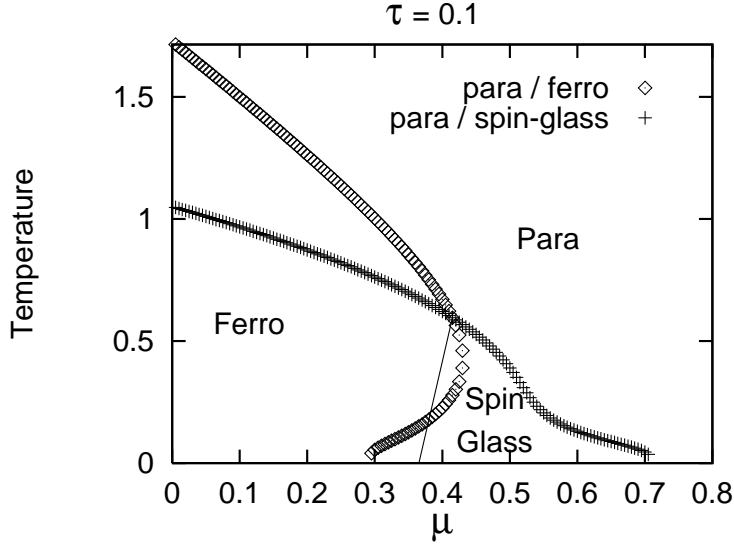,height=8cm}}
\caption{Phase diagram of the dilute magnet
model with frustration
($\lambda=0$, $\tau=0.1$).
The spin glass
phase exists below the percolating dilution $\mu_P =2^{-1/2}
\simeq 0.707$. The paramagnetic~/~spin glass phase
boundary inside the ferromagnetic phase is unphysical,
as well as the  paramagnetic~/~spin glass
boundary inside the spin glass phase. The solid line is
obtained from the calculation
in section~\ref{sec:dilutemodel} of the frontier
between the spin glass and ferromagnetic phases. This solution
is exact close to the critical point and we have continued
it to lower temperatures by an arbitrary linear behavior.
The exact zero temperature spin glass~/~ferromagnet phase boundary
is $\mu_0 \simeq 0.24191$.
}
\label{fig:para}
\end{figure}

\bigskip
\noindent
{\sl $\pm J$ model:}
The $\pm J$ model is recovered if $\mu=\tau=0$, with
a tricritical point at coordinates $p_t=1/2$,
$\lambda_t=\frac{1}{2} \left(1 - \frac{1}{\sqrt{2}}
\right)$~\cite{Carlson}.

\bigskip

\noindent
{\sl Dilute magnet with frustration:}
If $\lambda=0$ and $\tau \ne 0$
the relations (\ref{eq:tricritical}) become
\begin{eqnarray}
\label{eq:p-f}
&&\frac{2 p (1 - \mu)(1-u)(1 - \mu u p^2)}{1 - u p^2} = 1\\
&&\frac{2(1-\mu)p^2}{(1-u p^2)^2}
\left( (1-\mu)(1-u)^2 + \mu (1+u^2)
(1 - u p^2)^2 \right) = 1,
\label{eq:p-sg}
\end{eqnarray}
determining the paramagnetic phase boundary
shown on Fig.~\ref{fig:para}.
The frontier between the spin glass and
ferromagnetic phases will be examined
in Section~\ref{sec:reentrance}
by looking
for an instability in the first moment of the spin glass
solution.

The zero temperature limit of the paramagnetic/spin glass
phase boundary can be obtained by considering first the limit $p=1$
in Eq. (\ref{eq:p-sg}), and second the limit $u=1$. The order in
which the two limits are taken 
is imposed by the fact that $1-p \ll 1-u \ll 1$
at low temperatures because $\tau<1$.
One finds the limit to be $1-\mu = 1-1/\sqrt{2}$. 
As it is expected, this dilution is equal to the
percolation threshold of the Husimi cactus structure.

\section{Frontier between the spin glass and ferromagnetic phases}
\label{sec:reentrance}
\subsection{Method}

In order to determine the frontier between the spin glass
and ferromagnetic phases, we study the
instability of the spin glass solution with respect
to perturbations in the first moment. This involves
first calculating the spin glass solution
close to the tricritical line and next determining
whether this solution is stable with respect
to ferromagnetic fluctuations. Carlson {\it et
al.}~\cite{Carlson} performed this calculation
for the $\pm J$ model close to the tricritical point,
and shown 
the spin glass phase to be marginally reentrant
inside the ferromagnetic phase. By marginal, we mean
that the spin glass~/~ferromagnet phase boundary
has a quadratic behavior $\lambda_t - \lambda
\sim (T_t - T)^2$, which is specific to this model.
Other models (as the one we presently analyze)
have a linear behavior $\lambda_t - \lambda
\sim T_t - T$, with a positive (reentrant behavior)
or negative prefactor (non reentrant behavior).
The calculation follows Ref.~\cite{Carlson}
where the $\pm J$ model was solved,
and is asymptotically exact close to
the tricritical point.
This is complemented by an exact determination of the
zero temperature phase boundaries which, to
our knowledge, has not appeared previously
in the literature even for the
$\pm J$ model.

We first determine the asymptotic spin glass solution close to the
tricritical line. The second moment to lowest order is
\begin{equation}
\label{eq:2moment}
\LL m^2 \RR = \frac{2p^2 \LL G_{y,z}^2 \RR -1}{4p_t^2 \LL
G_{y,z} G_{z,y} H \RR_t}
,
\end{equation}
with
$
H = (p^2 \theta_y \theta_z - u)/
(1 - u p^2 \theta_y \theta_z)
$,
and the subscript ``t'' denoting a quantity evaluated
on the tricritical line.
We next consider a perturbation in the first moment of the
spin glass solution.
The recursion of the first moment is
$
\LL m \RR_{n+1} = \kappa \LL m \RR_n$
, with
$
\kappa=2p \LL G_{y,z} \RR -2p_t \LL m^2 \RR \LL G_{y,z} H 
\RR_t$
to order $(T_t - T)$.
If $\kappa < 1$ the spin glass phase
is stable and otherwise the ferromagnetic
phase is stable.

\subsection{$\pm J$ model
with an additional short-range coupling
$\tau$ -- Fixed $\mu$; ($\lambda$, T) phase diagram}
\label{sec:pmJ}

In the small-$\tau$ limit,
it is  straightforward to show that
\begin{enumerate}
\item{The tricritical point can be determined in an
expansion in $u_t$:
$
\label{eq:triperturb}
p_t=1 - 2 \lambda_t
=\frac{1}{\sqrt{2}}(1+\frac{u_t}{4})
$.
}
\item{The slope of the spin glass~/~ferromagnet
phase boundary at the tricritical point is
\begin{equation}
\label{eq:dk-dT}
\left. \frac{d \kappa}{d T} \right|_t = 
\frac{u_t}{4 T_t^0} \left( 1 - \frac{5 \sqrt{2}}{2 T_t^0}
\right) \simeq -0.467 \, u_t \mbox{ } < 0
,
\end{equation}
with $T_t^0 = 1/ \tanh^{-1}{(1/\sqrt{2})}
\simeq 1.135$ the tricritical point temperature with
$\tau=0$.
}
\end{enumerate}
>From what we deduce that the spin glass
phase does not reenter inside the ferromagnetic phase
close to the tricritical point.

\begin{figure}
\centerline{\psfig{file=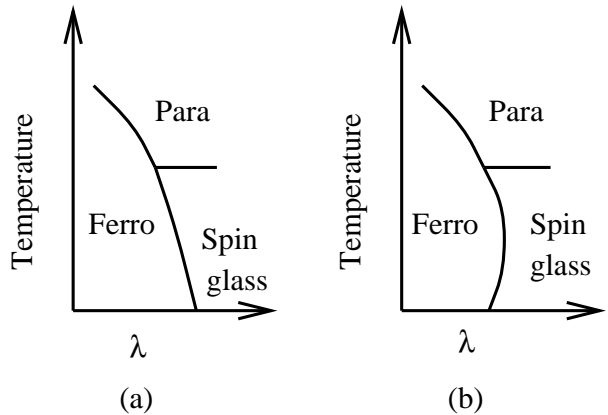,width=8cm}}
\caption{Possible shapes of the spin glass~/~ferromagnet
phase boundary in the $\pm J$
model with a small antiferromagnetic coupling $\tau$
corresponding to a spin glass phase not
reentrant inside the ferromagnetic phase asymptotically
close to the tricritical point.
This implies
two possible behaviors:
(a): no reentrance at any temperature (b): no reentrance
close to the tricritical point but reentrance at lower
temperatures. We prove that (b) is the correct
behavior in a zero temperature exact solution.
}
\label{fig:shape}
\end{figure}

Notice that $\left. d \kappa / d T \right|_t$
in Eq. (\ref{eq:dk-dT}) vanishes if $\tau=0$.
This is because reentrance is marginal in the $\pm J$
Bethe lattice spin glass~\cite{Carlson} and can
therefore not be obtained from an expansion to
first order in $T_t-T$.
We have evaluated numerically the coefficient
$\kappa$ with a finite $\tau$ and a finite $\mu$.
As $\mu$ is increased above a critical value,
the transition changes from reentrant
to non reentrant.

We now derive the exact spin glass solution in the
zero temperature limit, which
allows to discriminate rigorously between the two
behaviors on Fig.~\ref{fig:shape} (a) and~\ref{fig:shape}
(b).
We look for the zero temperature
fixed point spin glass
and ferromagnetic solutions $P^*(m)$ under the form
\begin{equation}
\label{eq:P*}
P^*(m) = \frac{x+y}{2} \delta(m-1) + (1-x) \delta(m)
+ \frac{x-y}{2} \delta(m+1)
,
\end{equation}
with $x$ and $y$ the spin glass and ferromagnet
order parameters.
It turns out that the functional form of the
magnetization distribution (\ref{eq:P*})
is stable when it is iterated in the zero
temperature limit of (\ref{eq:recursion}) and
(\ref{eq:PM}). 
To determine $x$ and
$y$, we impose (\ref{eq:P*}) to be the
fixed point magnetization distribution.
The solution with a finite magnetization is 
$
x = (1 - 4 \lambda)(1 - 2 \lambda)
$, and
$y^2 = (1-4 \lambda)(1 - 8 \lambda)/
(1 - 2 \lambda)^2
$.
Imposing $y^2>0$ leads to
the intersection $\lambda_0=1/8=0.125$ of the
spin glass~/~ferromagnet phase boundary
and the zero temperature axis in the $(\lambda,T)$
plane.
$\lambda_0$ is independent
of the strength of the additional coupling $\tau$.
If $\tau=0$, the value $\lambda_t$ of $\lambda$ at
the tricritical point is $\lambda_t=\frac{1}{2}(1
 - \frac{1}{\sqrt{2}}) \simeq 0.146$~\cite{Carlson},
larger than $\lambda_0$. The zero temperature
solution is therefore consistent with the reentrant
behavior of the spin glass~/~ferromagnet phase boundary
of the $\pm J$ Bethe lattice spin glass~\cite{Carlson}.
As $\tau$ is increased, the tricritical point
$(\lambda_t(\tau), T_t(\tau))$ evolves with $\tau$
whereas the intersection of the
spin glass~/~ferromagnet phase boundary and the
zero temperature axis remains equal to $\lambda_0$.
Therefore, if $\tau$ is small, $\lambda_t(\tau)$
remains larger than $\lambda_0$. From what we
deduce the existence of a turning point in the
spin glass~/~ferromagnet phase boundary (Fig.~\ref{fig:shape} (b)):
the spin glass phase does not reenter close to the
tricritical point whereas it reenters at lower temperatures.
We believe this behavior to be generic of
spin glass models with frozen exchanges and
we conjecture that a similar behavior may be
obtained in finite dimensional models.

\subsection{Dilution with a short-range
frustration $\tau$ -- $\lambda=0$; ($\mu$, T) phase diagram}
\label{sec:dilutemodel}

A small-$\tau$ perturbation calculation leads to
$
p_t = 1 - \frac{1}{2} u_t
$,
$\mu_t = \frac{1}{2} - \frac{1}{2} u_t$, and
$\left. d \kappa / d T \right|_t
= \tau /2 T_t^2 \mbox{ } > 0$,
which proves that the spin glass phase reenters
in the ferromagnetic phase close to the
tricritical point in the limit
of small $\tau$.
We have shown on Fig.~\ref{fig:para} the behavior
of the spin glass~/~ferromagnet phase boundary. This
phase boundary is exact only close to the tricritical
point, and we have continued it by an arbitrary straight line
at lower temperatures.
The reentrant behavior is confirmed by  zero
temperature exact results.
The paramagnetic~/~spin glass frontier 
intersects the zero temperature axis
at the percolation threshold, and the spin glass~/~ferromagnet 
frontier intersects the zero temperature axis
at $\mu_0$, the real root of $-10 \mu_0^3
+ 6 \mu_0^2 - 5 \mu_0 + 1 = 0$,
approximately $\mu_0 \simeq 0.24191$.
This confirms
the reentrant behavior of the spin glass transition
in the dilute magnet model with frustration.

\section{Diluting the spin liquid}
\label{sec:dil-spinliquid}
\begin{figure}
\centerline{\psfig{file=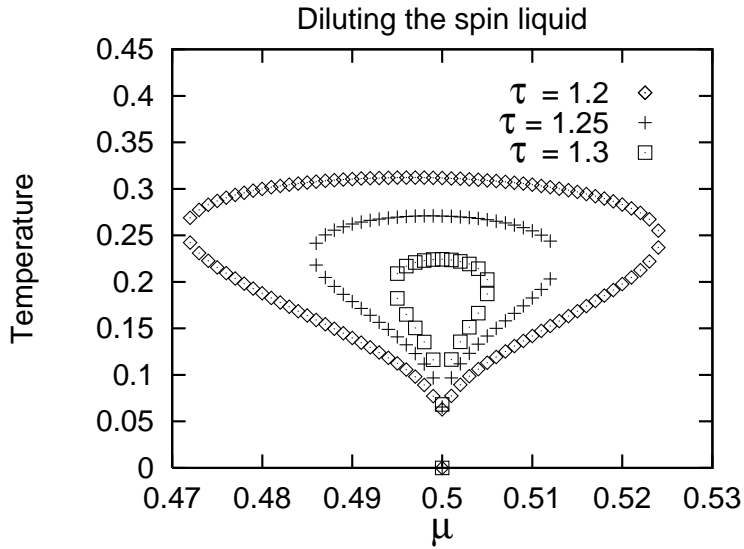,height=8cm}}
\caption{Boundary between the spin glass and
paramagnetic phases upon diluting the spin
liquid state of our model ($\tau>1$). The
spin glass phase is confined inside the
boundary shown for $\tau=1.2$ ($\Diamond$),
$\tau=1.25$ ($+$), and $\tau=1.3$ ($\Box$).
The spin glass boundary collapses onto
the point $\mu=1/2$ in the zero temperature limit.
This boundary behaves like $T \sim 1 / \ln{| \mu - 1/2|}$
around this singular point
The spin glass phase is favored upon increasing
the temperature (ordering by disorder -- see
section~\ref{sec:ordering-disorder}).
}
\label{fig:liquid}
\end{figure}

We now consider dilution
in the regime $\tau>1$, {\sl i.e.} when the pure system
is a spin liquid (see Fig.~\ref{fig:tau}). As it could
be expected, a ferromagnetic instability of
the spin liquid solution (Eq.~\ref{eq:p-f}) does
not exist. However, a spin glass instability
of the paramagnetic solution {\sl does} exist
upon diluting the system. Let us first consider
the zero temperature phases, in the limit
$1-u \ll 1-p \ll 1$ (since $\tau>1$).
The phase diagram at finite temperatures
is shown on Fig.~\ref{fig:liquid}.
A finite
temperature spin glass phase opens from the
point ($\mu=1/2$, $T=0$) as temperature
is increased from zero. The low
temperature phase boundary is
$
T = 4(1-\tau) / [\ln{(\mu-1/2)^2}]
$.
This provides a simple situation in which
diluting an Ising spin liquid results in an Ising spin glass
phase. The underlying ordering by disorder
mechanism~\cite{order}
is analyzed in section~\ref{sec:ordering-disorder}.

\section{Conclusion: diluting a frustrated magnet {\sl versus}
diluting a spin liquid}
We have shown the existence of a spin glass solution
upon diluting both the weakly frustrated magnet ($\tau <1$),
and the spin liquid ($\tau>1$).
We underline the differences in the physics in these
two regimes.

\subsection{Diluting the frustrated ferromagnet}

We believe
the generation of a spin glass phase upon
weakly frustrating a dilute magnet close to the
percolation threshold
to be due to the following: {\sl the strong diluted unfrustrated
magnet is already close to a spin glass}.
This can be seen on the example of square lattice dilute
magnets~\cite{Rammal}, where  dilution
removes sites in the ferromagnet up to the point
where the percolating cluster is a fractal object at
the percolation point. Since
the order of ramification of percolating
clusters is finite~\cite{ramification},
one can isolate large droplet-like objects~\cite{droplets}
from the remaining of the structure 
by cutting a finite number of bonds.
This results in large sets of
spins that can be reversed at a finite energy cost,
thus being responsible for the existence of quasi-degenerate
ground states separated by a large distance in phase space
(with different magnetizations~\cite{RM-fractal}),
and with barriers scaling like the logarithm of their
volume~\cite{Rammal}. 

The addition of frustration in dilute magnets 
close to the percolation threshold
turns the quasi spin glass order into a true one.
We have shown this explicitly in our model,
and a similar behavior was obtained in another
model~\cite{Georges-LeDoussal}.
We do not expect the low energy states of the spin
glass phase with a small frustration to be
very different from the droplet-like states
of the unfrustrated magnet.

The ``chaos and memory'' behavior of metallic
spin glasses was put forward
in Ref.~\cite{Bouchaud-droplets}, associated to the growth
of fractal droplets with a chaotic behavior in the sense
that droplets at a given temperature overlap weakly
with droplets at a different temperature.
We do not expect a chaotic behavior in our model
because the finite temperature droplet excitations
of the frustrated dilute magnet should be
obtained from reversing clusters of spins
in the unfrustrated magnet dilute lattice.

\subsection{Diluting the spin liquid: ordering by disorder}
\label{sec:ordering-disorder}
The mechanism for generating
a spin glass phase from the spin liquid
is different. 
The spin glass phase
originates from a balance between
the small-dilution regime in which {\sl dilution
suppresses the liquid behavior in favor
of spin-glass correlations}, and
a large-dilution regime in which {\sl dilution
suppresses spin glass correlations
by cutting the system into finite pieces}.
This is an order by disorder mechanism~\cite{order}:
thermal fluctuations favor a spin glass arrangement
and therefore reduce the phase space
dimensionality compared to the one of the
spin liquid state. 
Let us think
in terms of low temperature properties in
the large-$\tau$ limit. In this limit, the
neighboring spins coupled by the strong
antiferromagnetic exchange $\tau$
correlate antiferromagnetic, thus
leaving mainly two residual degrees of freedom
per bond $\tau$. We note $m_y$ and
$m_z$ the magnetization of these two spins,
and, for the sake of a qualitative argument,
assume $m_y=-m_z$ as a result of the strong
bond $\tau$. Let us assume the spins
at sites $y$ and $z$ to be frozen and look whether
freezing is relevant in the Bethe-Peierls
limit. We see from Eq. (\ref{eq:recursion})
that $m_x=0$ if the two ferromagnetic bonds
are present ({\sl i.e.} if the triangular
plaquette is frustrated, $\theta_y=\theta_z=1$).
In the unfrustrated plaquettes $\theta_y=0$, $\theta_z=1$,
or $\theta_y=1$, $\theta_z=0$, correlations
in the magnetization can propagate from one
generation to the other. The system
is cut into two pieces if $\theta_y=\theta_z=0$,
preventing correlations to propagate from
one generation to the other.
When the ferromagnetic bonds
are diluted, frustration is reduced since
the fraction of frustrated triangular plaquettes
$(1-\mu)^2$ decreases upon increasing the dilution
$\mu$. Decreasing frustration 
therefore decreases the short range liquid-like
correlations and favors a cooperative spin glass
arrangement.
In the large dilution limit, the exchanges are severely
depleted and a paramagnetic behavior is
restored since the system is cut into finite
pieces. In between these two regimes,
the unfrustred bond
configurations $\theta_y=1, \theta_z=0$
and $\theta_y=0$, $\theta_z=1$ 
with a weight $2 \mu(1-\mu)$ dominate
the physics and make a spin glass order
possible. 

\subsection{Concluding remarks}
Finally, we would like to compare the present work
to other approaches developed previously
in the literature, and mention some open questions.
The phase diagram with {\sl all} the bonds drawn from
the distribution (\ref{eq:bond}) was
studied by Aharony~\cite{Aharony},
Giri and Stephen~\cite{Giri} and Viana and
Bray~\cite{Viana-Bray}. These models share similarities
with the dilute fcc antiferromagnets studied
by de Seze~\cite{deSeze} and
Wengel, Henley and Zippelius~\cite{Henley3}.

Nieuwenhuizen and Nieuwenhuizen
and van Duin
studied the field theory of a model
of site-disordered magnet~\cite{Nieuwen1,Nieuwen2}.
One may also define a model similar to ours in
a finite dimension. As we conjectured, a phase
diagram similar to the one on Fig.~\ref{fig:shape} (b)
may be obtained. On the other hand, it may
be useful to investigate replica symmetry
breaking in Bethe-Peierls calculations.

Hierarchical lattices have been used previously
by Georges and Le Doussal~\cite{Georges-LeDoussal}
in relation with the renormalization group flow
along Nishimori's line, and by Gingras and
S{\o}rensen to study reentrance from a
paramagnetic to a ferromagnetic phase~\cite{Gingras}.
A model with frozen exchanges may be studied
on a finite dimensional hierarchical lattice,
which could be a first step in addressing
the phase diagram on Fig.~\ref{fig:shape}(b)
in a finite dimension.
This approach should probably rely on
a numerical iteration of the renormalization
equations similar to Ref.~\cite{Gingras}
while an analytic study was possible
in the present work.

Finally, ordering by disorder seems to be
a generic behavior of spin
liquids~\cite{Doucot,order,Quantum,Henley2,Simon}.
We found in the present work an ordering
by disorder resulting in a transition from
a paramagnetic to a spin glass ordering in an Ising
model. This may be viewed as a ``minimal''
ordering by disorder from a $Z_2$-symmetric
paramagnet to a spin glass because 
the Ising order parameter has the lowest possible
spin symmetry.

\section*{Acknowledgments}
The authors thank P. Chandra
and B. Dou\c{c}ot who introduced one of us (R.M.) to
Bethe-Peierls calculations. J. Souletie pointed
out to us the existence of a spin glass phase
upon diluting the spin liquid.
P. Simon pointed out to us Ref.~\cite{Henley2}. P.
Pujol pointed out to us
Refs.~\cite{Nishimori,Kitatani,Georges-LeDoussal}
and the absence of a Nishinori line argument.
The authors also acknowledge fruitful discussions
with A. Georges, P. Holdsworth,
P. Le Doussal, M. Gingras, J.M. Maillard,
H. Nishimori, and J. Villain.


\begin{thebibliography}{99}
\bibitem{note} A lot of other dilute ferromagnets
or antiferromagnets have been
studied experimentally since \protect~\cite{EuSrS}. See for
instance \protect~\cite{exp}, and references therein.
See also \protect~\cite{Henley3} and references therein.
\bibitem{Note} For instance, recasting the distribution
of antiferromagnetic couplings $\delta(J+\tau)$
under the form  $\exp{(A_{i,j}(\tau))}
Q_{i,j}(\tau)$ leads to
$A_{i,j}(-\tau)=+\infty$,
$A_{i,j}(\tau)=-\infty$ and $Q_{i,j}(\tau)
=Q_{i,j}(-\tau)=0$. Infinite
as well as vanishing quantities
are involved when we
expand averages
over the gauge group, making this calculation
uncontrolled.
\end{thebibliography}
\end{document}